\documentclass[prd,aps,showpacs,preprintnumbers,amssymb]{revtex4}
\usepackage{axodraw}
\usepackage{color}
\usepackage{epsf}

\def\e3p{$\eta \rightarrow 3 \pi$}

\begin{document}
\title{%
\hfill{\normalsize\vbox{%
\hbox{}
 }}\\
{Naturalness in a simple two Higgs doublet model}}
\author{Renata Jora
$^{\it \bf a}$~\footnote[1]{Email:
 rjora@theory.nipne.ro}}
\author{Salah Nasri$^{\it \bf b}$~\footnote[2]{Email:
 nasri.salah@gmail.com}}
\author{Joseph Schechter
 $^{\it \bf c}$~\footnote[3]{Email:
 schechte@phy.syr.edu}}

\affiliation{$^{\bf \it a}$ National Institute of Physics and Nuclear Engineering PO Box MG-6, Bucharest-Magurele, Romania}
\affiliation{$^{\bf \it b}$ Department of Physics, College of Science, United Arab Emirates University, Al-Ain, UAE}

\affiliation{$^ {\bf \it c}$ Department of Physics,
 Syracuse University, Syracuse, NY 13244-1130, USA}

\date{\today}

\begin{abstract}
We study the implications of a criterion of naturalness for a simple  two Higgs doublet model in the context of the discovery of a
Higgs like particle with a mass at 125 GeV. This condition which measures the amount of fine-tuning further limits the parameter space of this particular model and together with
other phenomenological constraints lead to an allowed range of masses for the other neutral or charged Higgs bosons: H, $a^{\pm}$, $a^0$.
\end{abstract}
\pacs{11.30. Qc, 11.15 Ex, 12.15. Lk}
\maketitle

\section{Introduction}
Recent experimental data from the LHC \cite{LHC1}-\cite{CMS2} suggests that a Higgs like particle with a mass of 125-126 GeV has been found.
Although this particle is consistent with a standard model Higgs boson it would be interesting to explore the consequences of this discovery for various extensions
of the standard model. Among these one of the most natural is the two Higgs doublet model. Many authors \cite{Posch}-\cite{Geller} have studied the parameter space of this type of model for the three possible scenarios: the 125 GeV Higgs boson is the lightest scalar in the model, the heaviest or the pseudoscalar $a^0$.

In the present work we analyze a particular case of the two Higgs doublet models introduced in \cite{Jora}. We will study the quadratic divergences of the scalars involved and in connection to the latest experimental data. More exactly we suggest that a criterion of naturalness should be applied also to this class of models.

We start with a two Higgs doublet model discussed in \cite{Jora} with a tree level effective Higgs potential that
satisfies the requirement of $SU(2)_L\times SU(2)_R$ flavor invariance together with parity and charge conjugation invariances.
We denote the two Higgs doublets by $\Phi$ and $\Psi$ where,
\begin{eqnarray}
\Phi=
\left[
\begin{array}{c}
i\pi^+\\
\frac{\sigma-i\pi_0}{\sqrt{2}}
\end{array}
\right],
\hspace{1cm}
\Psi=
\left[
\begin{array}{c}
-ia^+\\
\frac{\eta+ia_0}{\sqrt{2}}
\end{array}
\right].
\label{doubl54}
\end{eqnarray}

One can make three invariants:
\begin{eqnarray}
&&I_1=\sigma^2+\pi^2
\nonumber\\
&&I_2=\eta^2+a^2
\nonumber\\
&&I_3=\sigma\eta-\pi a
\label{inv5454}
\end{eqnarray}

Then the tree level potential can be written as:
\begin{eqnarray}
V=
\frac{\alpha_1}{2}I_1+\frac{\alpha_2}{2}I_2+\frac{\alpha_3}{4}I_1^2+\frac{\alpha_4}{4}I_2^2+\frac{\alpha_5}{4}I_3^2
+\frac{\alpha_6}{4}I_1I_2.
\label{oneloop657}
\end{eqnarray}

Since we consider the doublet $\Psi$ to have reversed parity with respect to $\Phi$ the potential does not contain a term linear in $I_3$ due to parity invariance.

For a reasonable range of parameters the potential admits a minimum with $\langle\sigma\rangle\neq 0$ and $\langle\eta\rangle=0$. (See\cite{Jora} for details. Note that we also slightly change the notation for the $\alpha_i$ to agree with the standard model two Higgs doublets).
The minimum condition reads:
\begin{eqnarray}
\alpha_1+\alpha_3v^2=0,
\label{min6767}
\end{eqnarray}

whereas the scalar masses are simply:
\begin{eqnarray}
&&m_{\sigma}^2=2\alpha_3v^2
\nonumber\\
&&m_{\eta}^2=\alpha_2+\frac{\alpha_5+\alpha_6}{2}v^2
\nonumber\\
&&m_{a_0}^2=m_{a^{\pm}}^2=\alpha_2+\frac{\alpha_6}{2}v^2.
\label{mass4353}
\end{eqnarray}

From the Higgs mass and the minimum  condition one can determine the two parameters $\alpha_1$ and $\alpha_3$. The masses of the other
scalars depend on the three unknown parameters $\alpha_2$, $\alpha_5$ and $\alpha_6$.
Assuming that the lightest Higgs coincides with the scalar discovered by Atlas and CMS with a mass $m_h=125-126$ GeV, except for some lower bounds we have little
 experimental information regarding $\eta$ and the $a$'s.

 In the present model one can add two lower limits stemming from the well known experimental knowledge on the Z width,
 \begin{eqnarray}
&&m_a>\frac{m_Z}{2}
 \nonumber\\
 &&m_a+m_{\eta}>m_Z.
 \label{lowlimit546}
 \end{eqnarray}

since the decays of Z to $a^+ +a^-$ and to $a^0+\eta$ are kinematically prohibited.

\section{Masses and couplings}

  We adopt the criterion of the cancellation of the quadratic divergences, the analogue of the Veltman condition \cite{Veltman} for the standard model.
  Thus we will ask that the masses and couplings are such that the quadratic divergences to all scalar masses in our model cancel. The corresponding set of conditions was derived
  by Newton and Wu \cite{Newton} for the most general two Higgs doublet model. Applied to our case this leads to two constraints. These are:
  \begin{eqnarray}
 &&-12m_t^2+3m_Z^2+6m_W^2+3m_h^2+(2\alpha_6+\frac{\alpha_5}{2})v^2=0
  \nonumber\\
  &&3m_z^2+6m_W^2+(6\alpha_4+2\alpha_6+\frac{\alpha_5}{2})v^2=0
  \label{setofcondfg454}
  \end{eqnarray}

 From these it is straightforward to determine:
 \begin{eqnarray}
6\alpha_4=3m_h^2-12m_t^2.
\label{res5454}
\end{eqnarray}

The latest experimental data suggest \cite{LHC1}-\cite{CMS2} that the actual mass of the Higgs boson is around 125-126 Gev. Thus the constraint in Eq (\ref{res5454}) would lead to $\alpha_4<0$ which is unacceptable
from the point of view of the vacuum structure.

A possible interesting way out is to generalize our simple two Higgs doublet so as to admit a vev different from zero also for the $\eta$.  For that we assume that the model is still parity
and charge conjugation invariant but that the vacuum spontaneously breaks the parity invariance.
In the situation when $\langle\sigma\rangle=v_1$, $\langle\eta\rangle=v_2$ the minimum equations for the potential become:

\begin{eqnarray}
&&\alpha_1+\alpha_3 v_1^2+(\frac{\alpha_5+\alpha_6}{2})v_2^2=0
\nonumber\\
&&\alpha_2+\alpha_4 v_2^2+(\frac{\alpha_5+\alpha_6}{2})v_1^2=0.
\label{min76868}
\end{eqnarray}

If we denote,
\begin{eqnarray}
&&\tilde{\sigma}=\sigma-v_1
\nonumber\\
&&\tilde{\eta}=\eta-v_2
\label{den6776}
\end{eqnarray}

then the mass eigenstates are obtained through the transformation:
\begin{eqnarray}
\left[
\begin{array}{c}
\tilde{\sigma}\\
\tilde{\eta}
\end{array}
\right]
=
\left[
\begin{array}{cc}
\cos{\alpha}&\sin{\alpha}\\
-\sin{\alpha}&\cos{\alpha}
\end{array}
\right]
\left[
\begin{array}{c}
h\\
H
\end{array}
\right],
\label{def543354}
\end{eqnarray}

where,
\begin{eqnarray}
\tan{2\alpha}=
\frac{(\alpha_5+\alpha_6)v_1v_2}{2(\alpha_3v_1^2-\alpha_4v_2^2)}.
\label{res4343}
\end{eqnarray}

We define as usual $\frac{v_2}{v_1}=\tan{\beta}$ where $v_1^2+v_2^2=v^2$.
Then the mass spectrum can be easily derived and we deduce,
\begin{eqnarray}
&&m_{h}^2+m_{H}^2=2v^2[\alpha_3 \cos^2(\beta)+\alpha_4 \sin^2(\beta)]
\nonumber\\
&&m_{h}^2 m_{H}^2=v^4[4\alpha_3 \alpha_4-(\alpha_5+\alpha_6)^2]\sin^2(\beta)\cos^2(\beta).
\label{eq3232}
\end{eqnarray}
and,

\begin{eqnarray}
m^2_{a^0,a^{\pm}}=-\frac{\alpha_{5}}{2}v^2.
\label{somm4353}
\end{eqnarray}

Since we still preserve the parity invariance at the level of the Lagrangian the fermion couple only to first Higgs doublet (type I Higgs doublet model).
For this case the Newton-Wu conditions \cite{Newton}, \cite{Ma} of cancellation of quadratic divergences read (we keep only the top and bottom quarks):
\begin{eqnarray}
&&3m_Z^2+6m_W^2+(6\alpha_3+2\alpha_6+\frac{\alpha_5}{2})v^2=12\frac{m_t^2}{\cos^2(\beta)}+12\frac{m_b^2}{\cos^2(\beta)}
\nonumber\\
&&3m_Z^2+6m_W^2+(6\alpha_4+2\alpha_6+\frac{\alpha_5}{2})v^2=0
\label{newcond6565}
\end{eqnarray}

The couplings of the Higgs with the top and bottom quarks in our model are:
\begin{eqnarray}
&&(h{\bar t}t)=(h{\bar t}t)_{SM}\frac{\cos(\alpha)}{\cos(\beta)}
\nonumber\\
&&(h{\bar b}b)=(h{\bar b}b)_{SM}\frac{\cos(\alpha)}{\cos(\beta)}
\label{res4443}
\end{eqnarray}

whereas the coupling of the Higgs with the gauge bosons W and Z read:
\begin{eqnarray}
&&(h W W)=(h W W)_{SM}\cos(\alpha+\beta)
\nonumber\\
&&(h Z Z)=(h Z Z)_{SM}\cos(\alpha+\beta)
\label{res444353}
\end{eqnarray}

From these one can compute the two photon decay rate of the Higgs boson \cite{Marciano}:
\begin{eqnarray}
\Gamma_{h\rightarrow \gamma\gamma}=\Gamma_{h\rightarrow \gamma\gamma}^{SM}
\frac{|8.35\cos(\alpha+\beta)-1.84\frac{\cos(\alpha)}{\cos(\beta)}|^2}
{|8.35-1.84|^2}.
\label{diphoy7676}
\end{eqnarray}

Here the value of $m_h=125.9$ GeV was used and 8.35 and -1.84 are the W and top loop contributions in the standard model.

\section{Discussion}

The model contains seven parameters. Two of them can be eliminated from the minimum equations. This leaves us with five parameters $\alpha_3$, $\alpha_4$, $\alpha_5$,
$\alpha_6$ and $\beta$.  We consider as input the mass of the Higgs boson $m_h=125.9$ GeV. Note that there are two possibilities: $m_h\leq m_H$ and $m_h> m_H$.  Using Eq. (\ref{eq3232}), Eq. (\ref{somm4353}) and Eq. (\ref{newcond6565}) we plot the square of the mass of the
Higgs boson H ($m_H^2$) in terms of $m_a^2$ for different values of $\sin^2(\beta)$ for which we consider increments of $0.1$.

It turns out that there are no positive  solutions for $m_H^2$ for values of $\sin^2(\beta)$ in the range  $0.1-0.5$. For $\sin^2(\beta)=0.6$ there are solutions only for
$m_a^2\leq 10000$ GeV. However we are looking for solutions with the diphoton decay rate of the Higgs boson equal or greater  than that of the standard model and with couplings $(h{\bar b}b)$
close to the standard model couplings. There are no solutions even close to our requirements for $\sin^2(\beta)=0.6$.
For $\sin^2(\beta)=0.7$ there are two reasonable sets of solutions for $m_H^2$ (see Fig.1) both for masses of the a's $m_a^2\leq 20000$ GeV. However only the second set of solutions (dashed line in Fig.1) give acceptable two diphoton decay rate for the Higgs boson as illustrated in Fig.2 and also correct couplings with the bottom quarks (see Fig.7 thick line).
For $\sin^2(\beta)=0.8$ the solutions for the masses $m_H^2$ are shown in Fig.3. The reasonable diphoton decay rates and bottom couplings correspond to the first set of solutions
in Fig.3 (thick line) and are displayed in Fig.4 and Fig.7 (dashed line).
The relevant graphs for $\sin^2(\beta)=0.9$ are given in Figs.5,6,7 (dotdashed line). Here again only the first set of solutions (thick line in Fig.5) gives the correct answers.

\begin{figure}
\begin{center}
\epsfxsize = 8cm
 \epsfbox{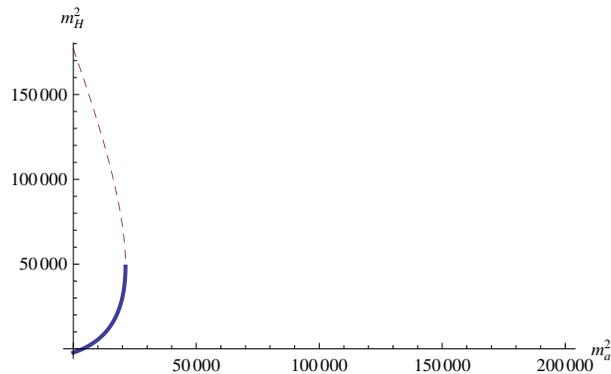}
\end{center}
\caption[]{%
The two solutions (thick line, dashed line) for $m_H^2$ as a function of $m_a^2$ (in $GeV^2$). Here $\sin^2(\beta)=0.7$.
}
\label{s081112}
\end{figure}

\begin{figure}
\begin{center}
\epsfxsize = 8cm
 \epsfbox{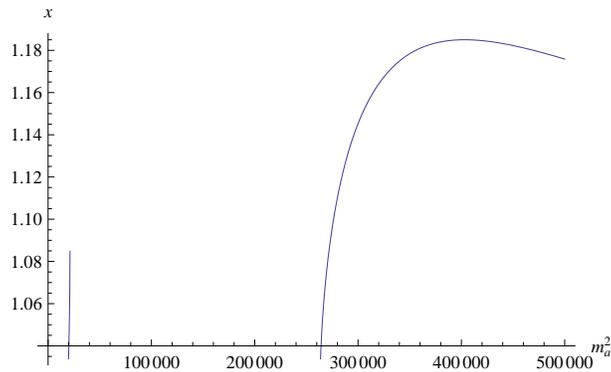}
\end{center}
\caption[]{%
The ratio $\frac{\Gamma_{h\rightarrow \gamma \gamma}}{(\Gamma_{h\rightarrow \gamma \gamma})_{SM}}=x$
 as a function of $m_a^2$ (in GeV). Here $\sin^2(\beta)=0.7$.
}
\label{gamma546}
\end{figure}

\begin{figure}
\begin{center}
\epsfxsize = 8cm
 \epsfbox{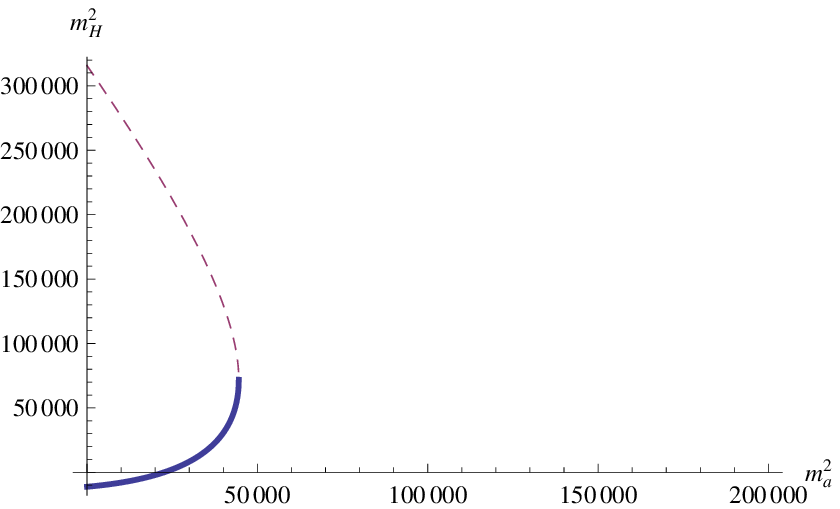}
\end{center}
\caption[]{%
The two solutions ( thick line, dashed line) for $m_H^2$ as a function of $m_a^2$ (in $GeV^2$). Here $\sin^2(\beta)=0.8$.
}
\label{s081332}
\end{figure}

\begin{figure}
\begin{center}
\epsfxsize = 8cm
 \epsfbox{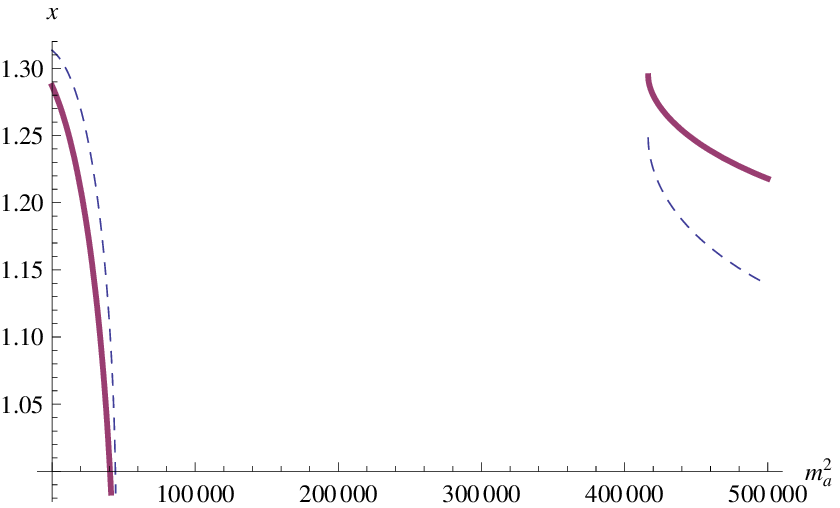}
\hskip 1cm

\end{center}
\caption[]{%
The ratio $\frac{\Gamma_{h\rightarrow \gamma \gamma}}{(\Gamma_{h\rightarrow \gamma \gamma})_{SM}}=x$
 as a function of $m_a^2$ for different choices of the sign of the angle $\alpha$. Here $\sin^2(\beta)=0.8$.
}
\label{s081552}
\end{figure}

\begin{figure}
\begin{center}
\epsfxsize = 8cm
 \epsfbox{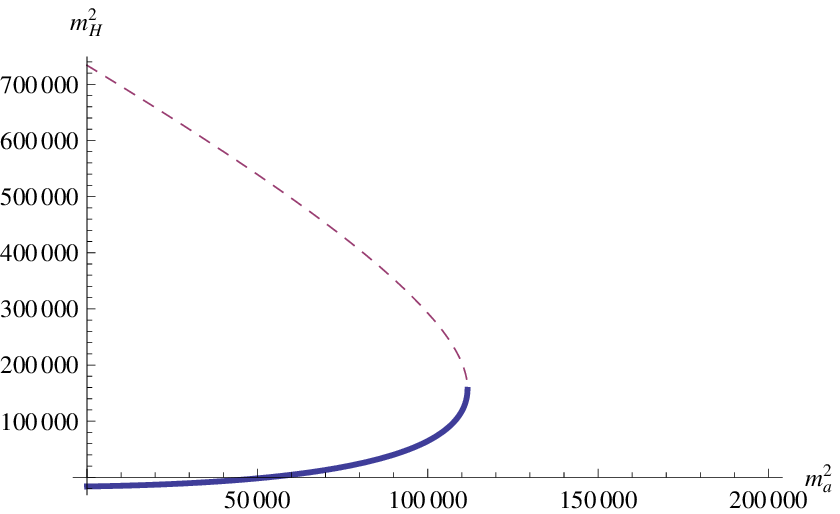}
\end{center}
\caption[]{%
The two solutions (thick line, dashed line) for $m_H^2$ as a function of $m_a^2$ (in $GeV^2$). Here $\sin^2(\beta)=0.9$.
}
\label{s081277}
\end{figure}

\begin{figure}
\begin{center}
\epsfxsize = 8cm
 \epsfbox{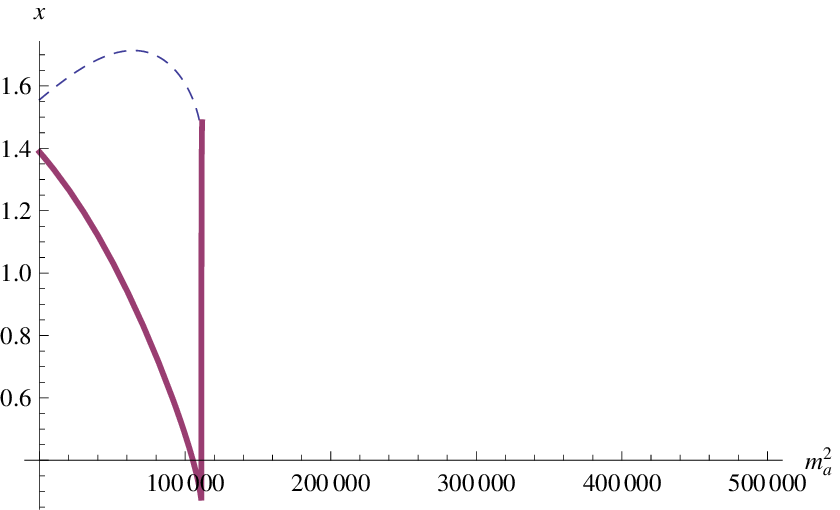}
\hskip 1cm

\end{center}
\caption[]{%
The ratio $\frac{\Gamma_{h\rightarrow \gamma \gamma}}{(\Gamma_{h\rightarrow \gamma \gamma})_{SM}}=x$
 as a function of $m_a^2$ for different choices of the sign of the angle $\alpha$. Here $\sin^2(\beta)=0.9$.
}
\label{s0813452}
\end{figure}

\begin{figure}
\begin{center}
\epsfxsize = 8cm
 \epsfbox{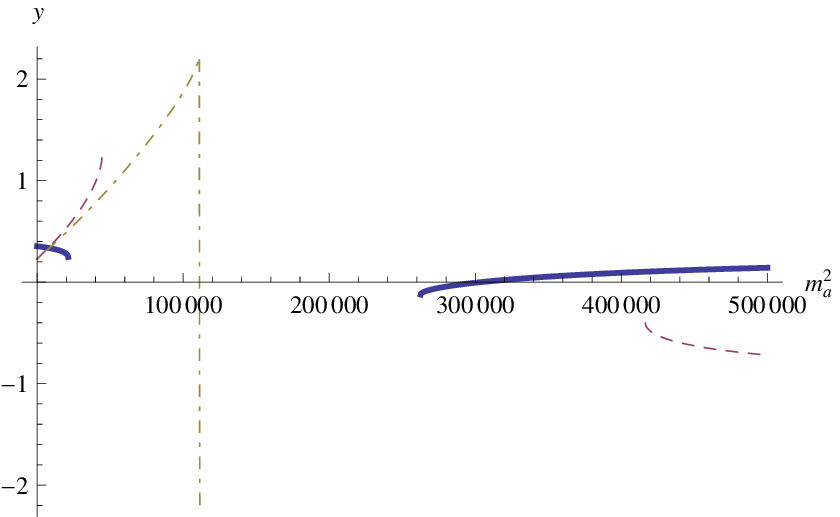}
\end{center}
\caption[]{%
The ratio $\frac{(hbb)}{(hbb)_{SM}}=y$
 as a function of $m_a^2$ (in GeV) for the three values of $sin^2(\beta)$: 0.7 (thick line),0.8 (dashed line),0.9(dotdashed line).
}
\label{b55213214}
\end{figure}

\section{Estimate of the masses}
The two Higgs doublet models have been discussed and analyzed extensively in the literature in connection to the LHC data \cite{Sher1},\cite{Sher2},\cite{Chang}.
It would be useful here rather then reiterate these efforts to apply some of these results to the naturalness problem. For this specific problem we will use the global fit for the parameters
$\alpha$ and $\tan(\beta)$  to the observed Higgs signal strength defined for all Higgs search channels at the LHC \cite{Chang}. The values of these parameters are then taken as inputs in Eqs. (\ref{res4343}), (\ref{eq3232}), (\ref{newcond6565}) together with the mass of the Higgs boson. The system of 5 equations leads to solutions for all the unknown parameters of the model:
$\alpha_3$, $\alpha_4$, $\alpha_5$, $\alpha_6$ and $m_H$.
The three scenarios displayed in Table I \cite{Chang} correspond to: 1) The mass of the lightest Higgs boson is 125-126 GeV; 2) The mass of the heaviest Higgs boson is 125-126 GeV ; 3) There are two resonances, h and $a^0$ with a mass around 125-126 GeV.

\begin{table}[htbp]
\begin{center}
\begin{tabular}{c|c|c|c}
\hline \hline
 ${\rm Masses}$&  I ($\alpha=1.38+\pi$, $\cot(\beta)=0.21$ )          & II($\alpha=-0.15+\pi$, $\cot(\beta)=0.17$)  & III($\alpha=-0.98+\pi$, $\cot(\beta)=1.37$)\\
 \hline
$m_H(m_h)$& $381\,{\rm GeV}$& $368\,{\rm GeV}$& $m_H^2<0$\\
$m_a$&  $132\,{\rm GeV}$ &  $129\,{\rm GeV}$& $305\, {\rm GeV}$\\
\hline
\end{tabular}
\end{center}
\caption{Masses of the Higgs bosons $H(h)$, $a^{\pm}$, $a^0$ for the three unconstrained scenarios.}
 \label{table22}
\end{table}

As it can be seen form Table I only the first scenario works as scenario II leads to an inconsistency ($m_h=368$ GeV) and scenario II leads to  a imaginary mass for the H boson.

We relax the condition (\ref{newcond6565}) and replace it by a new constraint which limits the amount of fine-tuning in this sector. First let us express the quadratic contribution to the scalars $\sigma$ and $\eta$ self energies before spontaneous symmetry breakdown:
\begin{eqnarray}
&&\delta m_{\sigma}^2=\frac{\Lambda^2}{32\pi^2 v^2}[3m_Z^2+6m_W^2+6m_W^2+(6\alpha_3+2\alpha_6+\frac{\alpha_5}{2})v^2-12\frac{m_t^2}{\cos^2(\beta)}-12\frac{m_b^2}{\cos^2(\beta)}]
\nonumber\\
&&\delta  m_{\eta}^2=\frac{\Lambda^2}{32\pi^2 v^2}[3m_Z^2+6m_W^2+(6\alpha_4+2\alpha_6+\frac{\alpha_5}{2})v^2]
\label{newcond656578}
\end{eqnarray}

Then we ask:
\begin{eqnarray}
&&\delta m_{\sigma}^2\leq 0.1 \alpha_1
\nonumber\\
&&\delta m_{\eta}^2\leq 0.1\alpha_2.
\label{newfine555}
\end{eqnarray}

Here $\alpha_1$ and $\alpha_2$ are the masses of the $\sigma$ and $\eta$ in the gauge eigenstate basis.
For large $\Lambda$ Eq. (\ref{newfine555}) approaches the condition of cancellation of the quadratic divergences such that we will study the implication for a $\Lambda$ relatively small,
 $\Lambda=10$ TeV.  For scenario I we plot the parameters $a_3$, $a_4$, $a_6$ (see Fig.8) and also the mass $m_a^2$ as a function of the allowed range for $m_H\leq 381$ GeV (Fig.9). It turns out that
 the mass $m_a$ is real only for $283\,{\rm GeV}\leq m_H\leq 381\,{\rm GeV}$ and increases from zero to $132$ GeV in this interval.
\begin{figure}
\begin{center}
\epsfxsize = 8cm
 \epsfbox{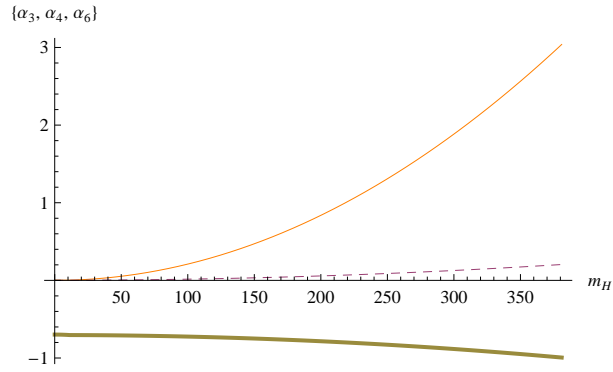}
\end{center}
\caption[]{%
Plot of the parameters $\alpha_3$ (dashed line), $\alpha_4$ (orange line) and $\alpha_6$ (thick line)as a function of the mass of the heavy Higgs boson $m_H$ in scenario I.
}
\label{ft5545}
\end{figure}

\begin{figure}
\begin{center}
\epsfxsize = 8cm
 \epsfbox{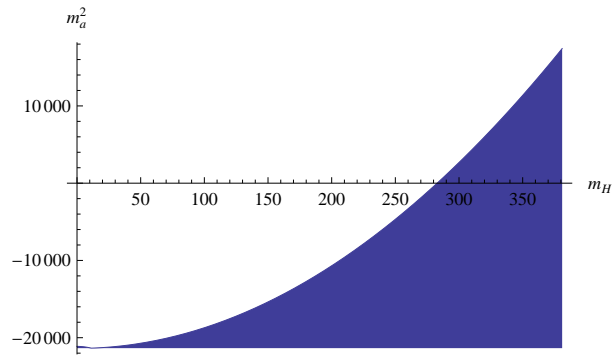}
\end{center}
\caption[]{%
Plot of the allowed  values for $m_a^2$ (grey region) as a function of the mass of the heavy Higgs boson $m_H$ in scenario I.
}
\label{ft5545}
\end{figure}

\section{Conclusions}

The naturalness criterion plays an important role in building beyond the standard model theories like supersymmetry, technicolor, extra dimensions, little Higgs etc.
Even if the two Higgs doublet model can be viewed as a lower effective limit of one of these theories or another one should still consider a measure of the fine-tuning that it is allowed
at least with respect to some scale at which new physics might intervene.

In the present work we consider for a particular type I two Higgs doublet model two cases: a) when the scale of new physics is high and  b) when the scale of new physics is relatively low. For case a) we apply the condition of cancellation of quadratic divergences and study this in conjunction with the Higgs diphoton decay rate and the $(h {\bar b}b)$ couplings. For case b) we require that the quadratic corrections to the scalar masses be relatively small with respect to the actual masses and analyze this in the context of more comprehensive phenomenological fits for the angle $\alpha$ and $\tan(\beta)$ taken from the literature \cite{Chang}. We thus estimate an acceptable interval for the masses of the other neutral and charged Higgs bosons: H, $a^{\pm}$, $a^0$. Depending on the set of conditions applied the range of masses can be larger or smaller.
Our conclusion is that our two Higgs doublet model can both be natural and in agreement with the latest experimental data.

\section*{Acknowledgments} \vskip -.5cm

The work of R. J. was supported by PN 09370102/2009.  The work of J. S. was supported in part by US DOE  under Contract No. DE-FG-02-85ER 40231.


\begin{thebibliography}{15}
\bibitem{LHC1} ATLAS Collaboration, Phys. Lett. {\bf B} 710, 49-66 (2012).
\bibitem{LHC2} ATLAS Collaboration, Phys. Lett. {\bf B} 716, 1 (2012).
\bibitem{CMS1} CMS Collaboration, Phys. Lett {\bf B} 710, 26 (2012).
\bibitem{CMS2} CMS Collaboration,  Phys. Lett {\bf B} 716, 30 (2012).
\bibitem{Posch} A. P. Posch, Phys. Lett. {\bf B} 696, 447 (2011).
\bibitem{Sher1} P. M. Ferreira, R. Santos, M. Sher, J. P. Silva, arXiv:1201.0019 (2012).
\bibitem{Sher2} P. M. Ferreira, R. Santos, M. Sher, J. P. Silva, arXiv:1112.3277 (2011).
\bibitem{Chien} C.-Y. Chen, S. Dawson, arXiv:1301.0309 (2013).
\bibitem{G1} B. Grzadkowski, P. Osland, Phys. Rev. D {\bf 82}, 125026 (2010), arXiv:0910.4068.
\bibitem{G2} B. Grzadkowski, P. Osland, Fortsch. Phys. 59, 1041-1045 (2011), arXiv:1012.0703.
\bibitem{G3} B. Grzadkowski, P. Osland, J. Phys. Conf. Ser. 259, 012055 (2010), arXiv:1012.2201.
\bibitem{Gerard} E. Cervero and J.-M. Gerard, Phys. Lett. B {\bf 712}, 255 (2012), arXiv:1202.1973.
\bibitem{Wang} L. Wang, X.-F. Han, JHEP 1205, 088 (2012), arXiv: 1203.4477.
\bibitem{Drozd} A. Drozd, B. Grzadkowski, J. F. Gunion, Y. Jiang, arXiv:1211.3580 (2012).
\bibitem{Haber} P. M. Fereira, H. F. Haber, R. Santos, J. P. Silva, arXiv:1211.3131 (2012).
\bibitem{Alves} B. D. S. M. Alves, P. J. Fox, N. Weiner, arXiv:1207.6499.
\bibitem{Chang} S. Chang, S. K. Kang, J.Lee, K. Y. Lee, S. C. Park, J. Song, arXiv:1210.3439 (2012).
\bibitem{Geller} S. Bar-Shalom, M. Geller, S. Nandi, A. Soni, arXiv:1208.3195 (2012).
\bibitem{Jora} R. Jora, S. M. Moussa, S. Nasri, J. Schechter and M. N. Shahid, Int. J. Mod. Phys. A 23, 5159 (2008), arXiv:0805.0293.
\bibitem{Veltman}M. J. G Veltman, Acta Phys. Polon. B 12, 437 (1981).
\bibitem{Newton} C. Newton and T. T. Wu, Zeitschrift fur Physik, 62, 253-263, 1994.
\bibitem{Ma} E. Ma, arXiv:hep-ph/0101355 (2001).
\bibitem{Marciano} W. J. Marciano, C. Zhang and S. Willenbrock, Phys. Rev. D {\bf 85}, 013002 (2012); arXiv:1109.5304.

\end{thebibliography}
\end{document}